# Seed-based information retrieval in networks of research publications: Evaluation of direct citations, bibliographic coupling, co-citations and PubMed related article score

Peter Sjögårde[a,b,x], Per Ahlgren [c]

[a]Health Informatics Centre, Department of Learning, Informatics, Management and Ethics, Karolinska Institutet, Stockholm, Sweden
[b]University library, Karolinska Institutet, Stockholm, Sweden
[c]Department of Statistics, Uppsala University, Uppsala, Sweden

ORCID:
[x]https://orcid.org/0000-0003-4442-1360
[y]https://orcid.org/0000-0003-0229-3073

Email: peter.sjogarde@ki.se; per.ahlgren@uu.se

Corresponding author: Peter Sjögårde, University Library, Karolinska Institutet, 17177 Stockholm, Sweden

## Abstract

In this contribution, we deal with seed-based information retrieval in networks of research publications. Using systematic reviews as a baseline, and publication data from the NIH Open Citation Collection, we compare the performance of the three citation-based approaches direct citation, co-citation, and bibliographic coupling with respect to recall and precision measures. In addition, we include the PubMed Related Article score as well as combined approaches in the comparison. We also provide a fairly comprehensive review of earlier research in which citation relations have been used for information retrieval purposes. The results show an advantage for co-citation over bibliographic coupling and direct citation. However, combining the three approaches outperforms the exclusive use of co-citation in the study. The results further indicate, in line with previous research, that combining citation-based approaches with textual approaches enhances the performance of seed-based information retrieval. The results from the study may guide approaches combining citation-based and textual approaches in their choice of citation similarity measures. We suggest that future research use more structured approaches to evaluate methods for seed-based retrieval of publications, including comparative approaches as well as the elaboration of common data sets and baselines for evaluation.

## Introduction

An essential principle in the research system is to build new research upon previous research and to contextualize new research using references in publications to previous research (de Solla Price, 1965; Tahamtan & Bornmann, 2022). Information searching is essential in this process. One important way for researchers to find publications reporting on previous research is by the use of information retrieval systems. Traditionally such systems retrieve and rank publications of relevance to a search query. However, this approach is sometimes associated with problems, such as being time consuming, having low recall, i.e., a low proportion of the relevant publications is retrieved, or low precision, i.e., a low proportion of the retrieved publications is relevant (Bramer et al., 2016; Gusenbauer & Haddaway, 2020).

Within the search query-based approach, one or more publications that are already known to be relevant may be used as seeds, and publications related to the seeds can subsequently be retrieved. In this case, terms associated with the seeds are used to construct a search query. Seed-based approaches of this type are supported by several information systems. An example is the list of "Similar articles" provided by PubMed. This list is based on the PubMed Related Article score (RA), which compares vocabulary in two publications (Lin & Wilbur, 2007).

The rationale of seed-based approaches, regardless of type, is that publications that are related to the seeds are likely to be considered as relevant to the information need of the user. A fundamentally different type of seed-based approach, relative to the one described in the preceding paragraph, is the use of citation relations. Ever since researchers started to use references in their publications, it has been possible for readers to find other relevant publications by following these references. The Web of Science was the first system making it possible also to go in the other direction, namely retrieving the publications citing a known publication (Garfield, 1955). Regardless of direction, the citation relation involved is direct citation (DC). Nowadays, many systems including citation information exist. Such systems may also make it possible to find publications that are indirectly related to seed publications through citations. Such publications can be related either by citing the same publication(s) (Fano, 1956), called bibliographic coupling (BC) (Kessler, 1963), or by being cited by the same publication(s), called co-citations (CC) (Marshakova-Shaikevich, 1973; Small, 1973).

In recent time, several systems have emerged that use citation-relations for information retrieval, including Inciteful[i], ResearchRabbit[ii], LitMaps[iii] and Connected Papers.[iv] The idea to retrieve publications related to seeds by citation relations is not new. Still, there is a lack of studies comparing the performance of different methods. Considering earlier studies on the theme, we further believe that the problem of how to best elaborate seed-based systems needs to be studied more structured. In this study, we focus on the retrieval performance of simple citation-based approaches for seed-based search, believing that our work can form a baseline for future studies that implement more advanced approaches.

To the best of our knowledge, no previous study has compared the performance of DC, BC and CC for seed-based information retrieval. In this paper, we compare the performance of these approaches. In addition, we include the RA score as well as combined approaches in the comparison. In the next section, we report on related research focusing on seed-based approaches. Data are presented in the following section. The section "Methods" treats the citation relations we have used for seed-based information retrieval. This section is followed by the results. Implications for seed-based approaches are discussed in the last section, including recommendations for future research.

## Related research

In this section, we review a selection of earlier research in which citation relations have been used for information retrieval purposes. We conclude the section by identifying a knowledge gap, which we address in this paper.

About 20 years ago, Larsen (2002) proposed a citation search strategy that starts with the publications, and their references, retrieved by a subject search. Additional publications are then identified algorithmically based on retrieving publications that cite references in intersections of sets of references, where the references occur in the initially retrieved publications. The strategy, which was capable of retrieving a large number of publications, does not require the user to supply seed publications in advance.

Bernstam et al. (2006) studied the performance of eight algorithms with respect to their ability to retrieve important articles, defined as being included in a pre-existing bibliography of important literature in surgical oncology. Three of the algorithms were based on citation relations. The results showed that the citation-based algorithms outperformed noncitation-based algorithms at identifying important articles. More precisely, simple citation count and PageRank had the best performance.

In Lu et al. (2007), a collection of computer science publications was used to compare the retrieval effectiveness of citation-based and text-based similarity measures. Two variants of citation-based similarity estimation between two nodes (publications) in a citation network were used, both based on the local neighborhoods of the nodes. According to the authors, the key lesson from the comparison was the complementary nature of citation-based and text-based approaches.

To enrich PubMed with a new type of related article link based on citations within a single sentence (sentence level co-citations) was proposed by Tran et al. (2009). The results of the study showed that publications linked by sentence level co-citations are highly related, where different similarity metrics were used in the assessments. The results further showed that there was a quite low overlap between sentence level co-citation pairs and PubMed related article links.

Liang et al. (2011) proposed a citation-based measure of the relatedness between two publications. The measure takes into account the distance between nodes in a citation network, as well as three types of citation relations (comparable, based-on and general). The measure was compared to other relatedness measures, including bibliographic coupling and co-citation, regarding the ability to find relevant publications with the use of a seed publication. It turned out that the proposed measure was more effective than the other measures in finding relevant publications.

Ortuño et al. (2013) proposed a query expansion method based on the cited references of a publication. The author used a collection of full-text biomedical publications. The use of cited references in the full publication and in its various sections (like Introduction and Results) was compared. In the experiments, two benchmarks were employed: retrieving a publication using its references, and retrieving topic-related publications using references from a single publication. It was concluded that cited references allowed accurate retrieval of the citing publication and also of publications related to six biomedical topics defined by particular Medical Subject Headings terms (MeSH terms). The best performance was often obtained when using all cited references, even if the use of the references from Introduction and Discussion sections led to similarly good results.

Janssens & Gwinn (2015) conducted two studies with PubMed as data source. The authors aimed at reproducing the results of literature searches for sets of published meta-analyses by the use of co-citations. In both studies, and for a given meta-analysis, co-citation frequencies for publications with randomly selected known publications (i.e., publications included in the meta-analysis) were extracted, and articles with a score above a selection threshold were screened. In the second study, the method was extended by obtaining the direct citations for all publications retrieved by the co-citation approach. In both studies, a large median proportion of the studies included in the meta-analyses were retrieved, while a low median proportion publications were screened, relative to the number of publications screened for the original meta-analysis publications. Janssens et al. (2020) conducted a validation study with regard to the search methodology proposed in the earlier study. In the validation study, randomly selected, published systematic reviews and meta-analyses were used. The search methodology retrieved a median of 75% of the publications included in the systematic reviews and meta-analyses.

Steinert et al. (2015), in a study on scientific publication recommendations, proposed an approach in which the similarity, citation-based as well as text-based, between citing and cited publications is taken into account in order to eliminate irrelevant citations. The approach aimed at finding relevant publications relative to a seed publication. The evaluation, in which human raters evaluated the paper recommendations, showed that the proposed approach gave rise to good results, comparable to those generated by the collaborative filtering approach.

Like the studies by Janssens & Gwin (2015) and Janssens et al. (2020), the study by Belter (2016) aimed at reproducing the results of literature searches. The author proposed starting with a few seed publications and then identifying cited, citing, co-cited, and co-citing publications to retrieve additional publications that might potentially be related to the seeds. The publications retrieved by the approach were compared to those retrieved by 14 traditionally performed systematic reviews. The approach retrieved 74% of the studies included in these reviews and 90% of the studies it could be expected to retrieve. Further, the approach retrieved substantially fewer records overall than the reviews did and was performed with substantially less time and effort. In a follow-up study, Belter (2017) proposed a method to relevance rank the publications retrieved by the citation-based search approach of the earlier study. The method uses the number of citation overlaps–with respect to direct citations, bibliographic coupling and co-citation–that a publication shares with the seed papers as the relevance ranking for the publication. With regard to the results, publications with high relevance scores represented slightly more than a quarter of the publications retrieved by the method, and approximately one-tenth of the publications retrieved by the reviews, but included three-quarters of the studies included in those reviews.

Habib & Afzal (2017), in the context of publication recommendation, proposed an approach that extends the traditional bibliographic coupling by exploiting the proximity of in-text citations of bibliographically coupled publications. The in-text citations were clustered by a density-based algorithm. The results of the experiments, in which human raters were used, showed that the accuracy of the recommendations for the proposed approach increased compared to traditional bibliographic coupling and text-based approaches. Traditional bibliographic coupling was extended, in a publication recommendation context, also in the study by Habib & Afzal (2019). However, in this work the distribution of citations in logical sections in bibliographically coupled papers was exploited, and automated evaluation (using Jensen-Shannon divergence) was used. The results showed that the proposed approach performed better than traditional bibliographic coupling and content-based research publication recommendation.

Khan et al. (2018) proposed an extension of the co-citation approach that exploits in-text citation frequencies and in-text citation patterns of the co-cited publications in different logical sections of the citing publication. The extension was experimentally compared to the traditional co-citation approach and to citation proximity analysis. For the results, the proposed approach outperformed the two other approaches in most of the cases.

Using four journals from different disciplines, Colavizza et al. (2018) analyzed the similarity of article pairs that are co-cited at different levels: journal, article, section, paragraph, sentence, and bracket. Four different measures of similarity between pairs of co-cited articles were used, namely textual similarity, intellectual overlap, author overlap, and proximity in publication time. The results indicated that the similarity of pairs of articles increases monotonically with their co-citation level, regardless of similarity measure.

SciRide Finder was proposed by Volanakis & Krawczyk (2018). This tool offered a literature search strategy that focuses on Cited Statements: any sentence from a peer reviewed publication containing citations to other manuscripts, which the authors referred to as Primary Research. A SciRide Finder user query, like "Cas9 Block", retrieved Cited Statements, and each search result consisted of, alongside a Cited Statement, the title of the publication that the Cited Statement appears in and the Primary Research sources that the Cited Statement refers to. The authors demonstrated that Cited Statements can carry different information retrieval data to those found in titles, abstracts and full text of publications they refer to.

An interactive visual tool for scientific literature search was presented by Bascur et al. (2019). Given an initial set of retrieved publications, the tool scatters the publications into labeled clusters, where the clustering is based on direct citation relations between publications. The user can then gather clusters of interest, and the union of gathered clusters can be scattered into new clusters, some of which the user can gather. The number of clusters in each scatter iteration is set by the user. The tool further visualizes the clusters as bubbles in a packed bubble chart. The size of the bubbles reflects the number of publications in the clusters and the distance between the bubbles approximately reflects the number of citation relations between the clusters.

Eto (2019) proposed a co-citation search technique, which is graph-based publication retrieval on a co-citation network containing citation context information. The technique expands the scope of seed publications by iteratively spreading the co-citation relation in order to obtain relevant publications that are not identified by traditional co-citation searches. Co-citation contexts were taken into account in the study, which involved a set of graph-based algorithms to compute the similarity between seeds and other publications. The results showed that those proposed methods that used co-citation contexts (some of the proposed methods did not) tended to perform better than the used baselines, one of which was the traditional co-citation approach. Rodriguez-Prieto et al. (2019) also dealt with seed publications in a co-citation context. The authors proposed a statistical inference approach in which two co-cited publications are considered to be semantically related if they are co-cited more frequently than would be expected by pure chance. The probability of co-citation was obtained from a null model, which defined what was considered as pure chance. The approach could rank publications co-cited with a seed by minimizing the probabilities. The results showed that the approach could provide publications similar to a seed in terms of terminology, but also publications presenting other kinds of relations.

A term-based and citation network-based exploratory search system for COVID-19 was developed by Zerva et al. (2021). The system, which makes use of both direct and indirect citations, has an interactive search and navigation interface. A user evaluation with domain experts was conducted, and in general, most users were satisfied with the relevance and speed of the search results.

Using a published systematic review on software engineering, Ali & Tanveer (2022) compared the two citation databases Google Scholar and Scopus regarding effectiveness and usefulness. Primary studies included in the systematic review were searched in both databases, where the searches used citations to seed publications. The results for effectiveness indicated that Scopus was more effective than Google Scholar. For instance, a high proportion of the primary studies found by the Google Scholar search were also found in the Scopus search, and with higher precision. For usefulness, the authors concluded that Scopus is more transparent with regard to what is indexed in the database than Google Scholar, has metadata of higher quality and do not index, in contrast to Google Scholar, e.g., unpublished reports and teaching material.

Bascur et al. (2023) evaluated the performance of citation-based clusters for information retrieval tasks using 25 systematic reviews. A search process was simulated with a tree hierarchy of clusters and a cluster selection algorithm, which used the F-score of retrieving the publications in a cluster. The Boolean queries self-reported by the systematic reviews were, in order to serve as a reference, replicated by the authors. It was found that search performance of the citation-based clusters was highly variable and unpredictable and that the clusters work best for users that prefer recall over precision. It was also found that the clusters were able to complement query-based search by identifying additional relevant publications.

The previous studies treated in this section have suggested new solutions for seed-based information retrieval in an ad hoc manner. For example, it is not clear from the literature which citation-based approaches perform better than others. We suggest that the problem of how to best elaborate seed-based systems need to be addressed in a more structured way. In view of this suggestion, this work focuses on simple implementation of citation-based approaches for seed-based search, with the aim of elucidating the question which citation-based approach performs best in a seed-based information retrieval setting. We believe that this study can form a baseline for future studies that implement more complex approaches. However, complex methods are regularly less transparent, less efficient and their implementation is more costly in terms of expertise and computational resources. Therefore, they must prove to perform considerably better than approaches that are more simplistic.

## Data

Consider a researcher addressing a research question, *q*. The researcher is interested in all previous research related to *q*. Let us denote the corresponding set of publications as *P(q)*. Now assume that the researcher is already familiar with some of the publications in *P(q)*. We denote this proper subset of *P(q)* as *P(q)_s* and the remaining publications in *P(q)* as *P(q)_not-s*. The task for a seed-based approach is to find all publications in *P(q)_not-s* by using the publications in *P(q)_s* as seeds and retrieve publications related to these seeds. In this paper, we compare how well different seed-based approaches succeed in retrieving the publications in *P(q)_not-s* from the publications in *P(q)_s*.

To compare different approaches, we used a set of systematic reviews. We assume that each systematic review corresponds to a research question (*q*), and that the reference list of the systematic review contains the publications relevant to that research question, i.e., being a proxy for *P(q)*. We do of course recognize that this assumption is not entirely fulfilled, since there might be references in a systematic review with low relevance to the research question, or publications of high relevance that have not been included in the reference list. However, we consider the indicated approximation as good enough for comparisons of different approaches to retrieve relevant publications from seeds.

We used publication data from the NIH Open Citation Collection (NIH OCC) (iCite et al., 2024, April version 2024). NIH OCC is a citation index of PubMed indexed publications. We made no restrictions regarding publication types. We denote the set of NIH OCC publications as *P*. By using *P* we restrict the analysis to biomedicine. We made a random sample of 3,000 systematic reviews from the year 2022, operationalized as publications including the term "systematic review" in their title. We disregarded systematic reviews with less than 30 references, counting references that can be connected to another publication in *P* with the publication year between 2010 and 2021. We denote the set of selected systematic reviews as $R = \{r_1, ..., r_{3000}\}$, where *R* is a subset of *P*. For a systematic review $r_i$ in *R*, we consider the references of $r_i$ to publications from 2010-2021 as a proxy for the publications related to the research question of $r_i$. We denote this set as $P(r_i)$. Note that $P(r_i)$ is restricted to publications that are covered in the data source and can be reached by tracing citations. This restriction is intentional, because we limit our analysis to comparison of seed-based approaches and do not address coverage of data sources (for such analyses, see e.g. Z. Liang et al., 2021; Martín-Martín et al., 2021; Visser et al., 2021). The proportion of publications in PubMed having references (outgoing citation relations) in NIH OCC grows from about 88% for the publication year 2010 to about 92% for the publication year 2021. We randomly selected $n = 5$ publications from $P(r_i)$ to be used as seeds, and we denote this set as $P(r_i)\_seeds$ and the remaining publications in $P(r_i)$ as $P(r_i)\_not\-seeds$.

## Methods

Four different approaches were used to retrieve publications related to the seeds: three citation-based approaches, DC, BC and CC, and one text-based approach, RA. Furthermore, we tested two combined approaches: (1) the combination of the three citation-based approaches (DC-BC-CC), and (2) the combination of all four approaches (DC-BC-CC-RA).

The calculations of the relatedness scores using the different approaches are described in the following subsections.

There are different options to normalize citation-based approaches. For example, the bibliographic coupling strength between two publications can be normalized to the number of references in the publications. We did not perform such normalization, because we are interested in a comparison between basic implementations of the citation-based measures. We find basic knowledge about the performance of the different unnormalized and raw approaches to be essential before the elaboration of more complex approaches.

## DC

The DC score between a publication $p$ in $P$ and $P(r_i)\_seeds$ is the number of times $p$ is cited by any of the publications in $P(r_i)\_seeds$ plus the number of times $p$ cites any of the publications in $P(r_i)\_seeds$. Theoretically, the score of DC cannot be higher than the number of seeds in $P(r_i)\_seeds$. Note that we, in the calculation of DC scores relative to $P(r_i)\_seeds$, excluded certain publications in $P$: obviously $r_i$, but also the publications in $P(r_i)\_seeds$. Note that we have assumed that the publications in $P(r_i)\_seeds$ are already familiar to the hypothetical researcher. The exclusion operations were applied also to the approaches BC, CC and RA, and thereby to the approaches DC-BC-CC and DC-BC-CC-RA.

## BC

The BC score between $p$ and one of the publications in $P(r_i)\_seeds$ is the number of references within $P$ that they have in common. The BC score between $p$ and $P(r_i)\_seeds$ is the sum of the BC scores between $p$ and all of the publications in $P(r_i)\_seeds$. For efficiency reasons, we delimited the list of publications retrieved by BC to those publications having a BC score of at least 2.

## CC

The CC score between $p$ and one of the publications in $P(r_i)\_seeds$ is the number of times these two publications co-occur in the reference lists with respect to the publications in $P$. Note that we disregard $r_i$ in the calculation, since $r_i$ clearly co-cite each publication in $P(r_i)\_not\-seeds$, i.e. the publications we want to retrieve, and each publication in $P(r_i)\_seeds$, i.e. the seed publications for $r_i$. The CC score between $p$ and $P(r_i)\_seeds$ is the sum of the CC scores between $p$ and all of the publications in $P(r_i)\_seeds$. For efficiency reasons, we delimited the list of publications retrieved by CC to those publications having a CC score of at least 2.

## RA

The RA score compares the similarity of the bibliographic information in two publications. It is based on words in titles, abstracts, and Medical Subject Headings (MeSH). Words in the title are given more weight. Words are defined as "unbroken string of letters and numerals with at least one letter of the alphabet in it". Stemming and removal of stop words is performed. The algorithm also takes into account the (1) frequency of a term in the document, (2) length of the document, i.e. the total number of retrieved words in the document, and (3) total number of times the term occurs in other documents (Lin & Wilbur, 2007; PubMed, 2023). After calculations of term weights, the similarity of two documents is calculated as the sum of all weights of the terms they have in common.

The RA score between $p$ and $P(r_i)\_seeds$ is the sum of the RA scores between $p$ and all of the publications in $P(r_i)\_seeds$.

## DC_BC_CC

Combining the different approaches into one score is not trivial, because the different approaches result in different scales. BC generally retrieves many more publications than the other approaches. Furthermore, DC takes a value that cannot be higher than the number of seeds (at least not theoretically), while BC and CC may range to much higher values. We consider the question about how to combine different approaches to be an empirical question, which we do not address in this paper. For this study, we use a somewhat arbitrary approach based on experience.

For a publication $p$ with the scores $p\_DC$, $p\_BC$ and $p\_CC$, the $DC\_BC\_CC$ score between $p$ and $P(r_i)\_seeds$ is the sum of $\frac{p\_DC}{1}$, $\frac{p\_BC}{10}$ and $\frac{p\_CC}{10}$.

Some testing revealed that the maximum BC score for 100 random reviews was about 18 times higher than the maximum DC score. The maximum CC score was about 30 times higher than the maximum DC score. Considering these relationships, our way of combining the three approaches gives more weight to CC than to BC and less weight to DC in practice. We find this to be reasonable in view of an observed better performance of CC compared to BC and DC and of BC compared to DC.

## DC-BC-CC-RA

We consider also the question of how to combine citation-based approaches with textual-based approaches to be an empirical question. In this case we use an approach giving equal weights to the combination of DC, BC and CC and RA. This is done by retrieving the highest and lowest values from DC_BC_CC distribution and rescaling the vector of RA scores to this range.

The DC_BC_CC_RA score between a publication $p$ and $P(r_i)\_seeds$ is the sum of the $DC\_BC\_CC$ score between $p$ and $P(r_i)\_seeds$ and the rescaled $RA\_score$ between $p$ and $P(r_i)\_seeds$.

### Evaluation measures

To evaluate the results, we use the measures recall and precision. We present recall and precision values as percentages. For a systematic review, $r_i$, we calculate the number of publications in $P(r_i)\_not\text{-}seeds$ that have been retrieved using $P(r_i)\_seeds$ for each of the six approaches. We refer to this number as the number of *hits*. The *total recall* with respect to $r_i$ and a given approach is the number of hits for the approach divided by the number of publications in $P(r_i)\_not\text{-}seeds$. The *total precision* with respect to $r_i$ and a given approach is the number of hits divided by the total number of publications retrieved by the approach.

We also measure recall and precision, for a systematic review $r_i$ and a given approach, at different ranks in the list of retrieved publications. Recall@k is the number of hits among the top $k$ retrieved publications divided by the number of publications in $P(r_i)\_not\text{-}seeds$, whereas Precision@k is the number of hits among the top $k$ retrieved publications divided by $k$. In the case of ties (publications having the same score), we order the publications randomly.

## Results

Figure 1 shows the distributions of the number of retrieved publications over the 3000 systematic reviews as violine wrapped boxplots. BC resulted in the highest number of retrieved publications, if not considering the combined approaches. Combining the approaches resulted in slightly higher numbers of retrieved publications than for BC.

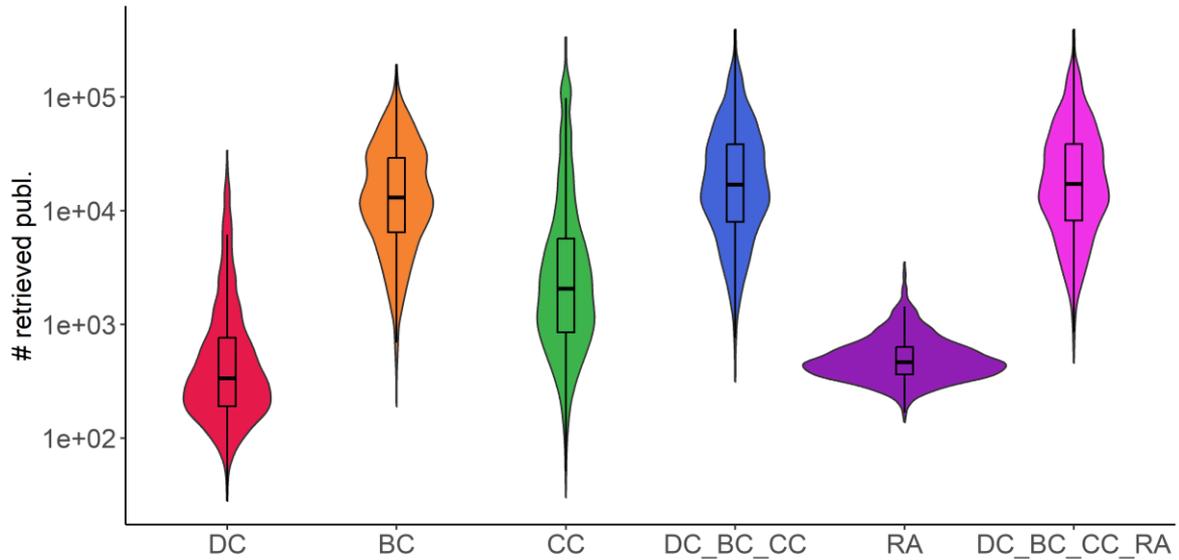

*Figure 1: Violine wrapped boxplots showing the number of retrieved publications by the different approaches. Log-10 scale on the y-axis.*

BC also resulted in the highest total recall when not considering the combined approaches (Figure 2). However, combining the three citation-based approaches (DC_BC_CC) led to markedly higher recall. Combining all approaches (DC_BC_CC_RA) resulted in about the same recall distribution as for DC_BC_CC. This indicates that RA does not add much value in terms of total recall.

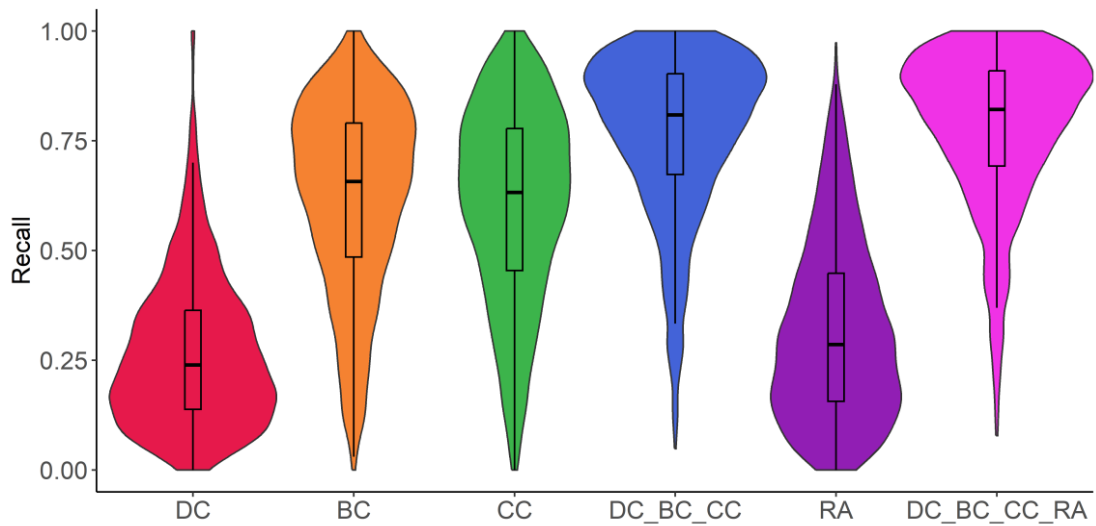

*Figure 2: Violine wrapped boxplots showing the total recall by the different approaches.*

Figure 3 reveals a low total precision for BC. The total precision is somewhat higher for DC, CC and RA. Considering all retrieved publications when calculating precision and recall shows a tradeoff between the two measures. The approaches with high total recall have low precision, while approaches with lower total recall have higher total precision.

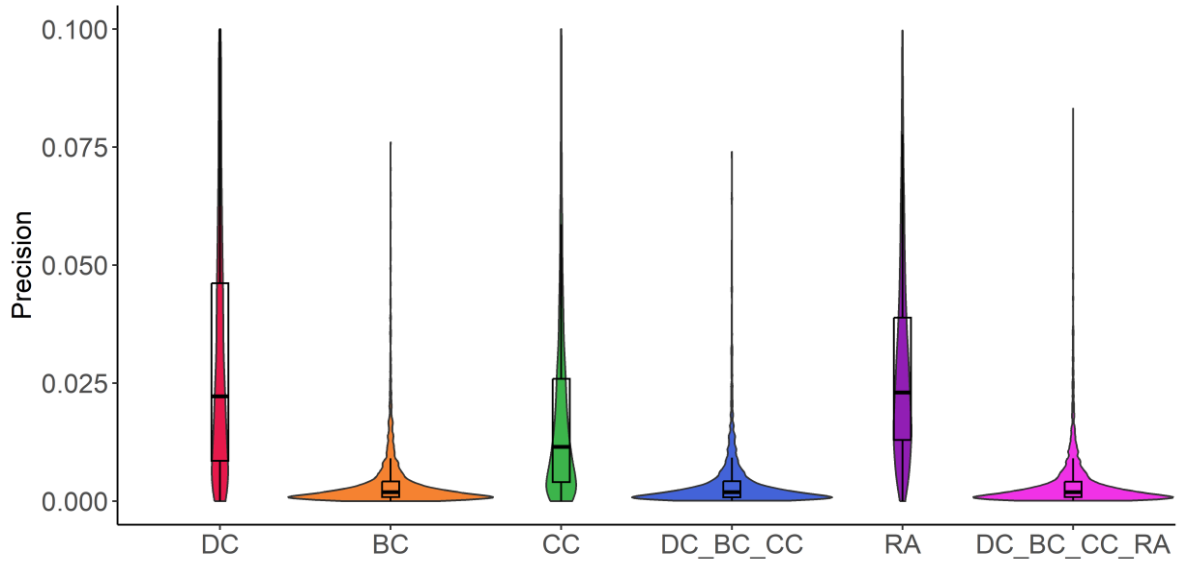

*Figure 3: Violine wrapped boxplots showing the total precision by the different approaches. Y-axis cut at 10.*

A user searching publications in an information retrieval system rarely browses through all of the retrieved publications. Therefore, the *Recall@k* measure is of high interest.

Figure 4 shows the average recall from $k = 1$ to $k = 100$. CC performs best when not considering the combined approaches. Among the 50 highest ranked publications by CC one can expect to retrieve about 15% of all publications in *P(r_i)_not-seeds*. RA is the second most successful approach retrieving about 12% of the publications in *P(r_i)_not-seeds* when $k = 50$. The DC and BC approaches perform similarly. However, the DC approach grows more slowly with increasing $k$, but DC performs better than BC when $k < 50$, whereas the opposite is the case when $k > 50$. Combining all approaches gives the highest recall when $k = 1$ to $k = 100$. However, combining only the citation-based measures (DC_BC_CC) gives almost as good results, especially at high values of $k$ (cf. the two corresponding confidence intervals).

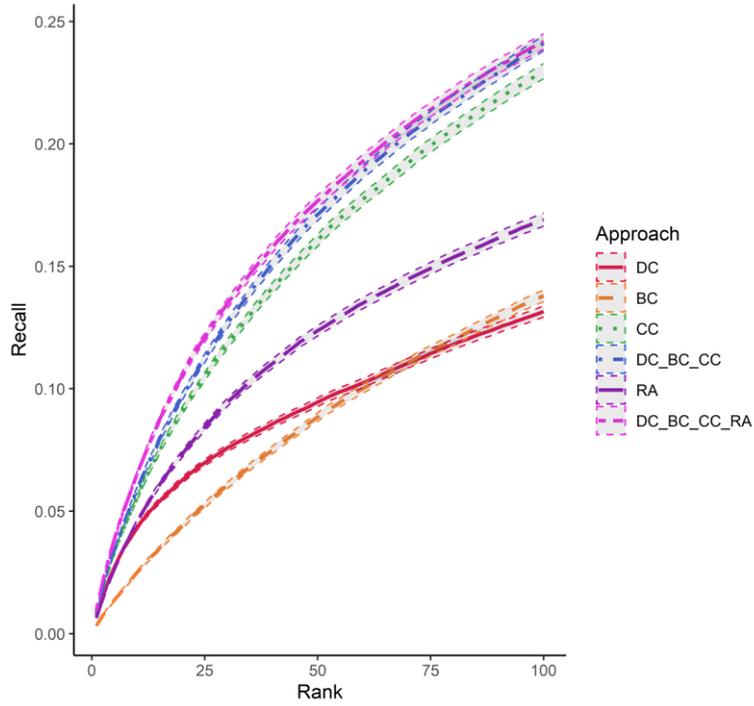

*Figure 4: Recall@k (k = 1, 2, ..., 100) of the approaches with a 95% confidence interval.*

Figure 5 shows the average precision from $k = 1$ to $k = 100$. For DC_BC_CC_RA the precision is around 0.17 when $k = 50$. It is only slightly less for DC_BC_CC and CC. For RA it is just above 0.1 at $k = 50$. BC and DC have a precision around 0.08 at $k = 50$. As in the recall case (Figure 4), DC performs better than BC when $k < 50$, whereas the opposite is the case when $k > 50$.

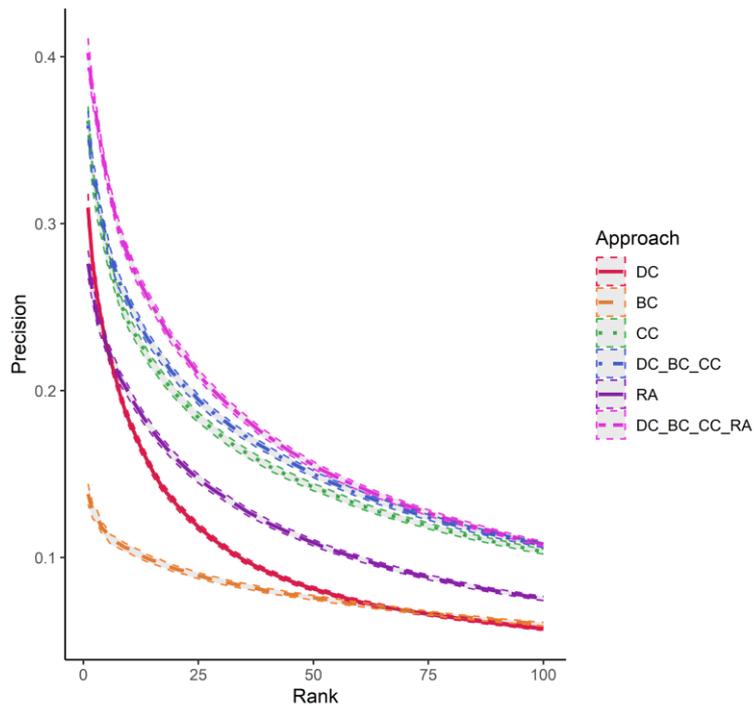

*Figure 5: Precision@k (k = 1, 2, ..., 100) of the approaches with a 95% confidence interval.*

For higher values of $k$, the value of combining the different citation-based measures increases (Figure 6). The DC_BC_CC approach retrieves around 50% of the publications in *P($r_i$)_not-seeds* at $k = 1000$. It may seem unlikely that a user browses through 1000 records in a search result. However, this is not unlikely when conducting a systematic review. Therefore, the performance at rather high values of $k$ may be of importance. Note that the recall using DC flattens after about 500 records. It is therefore likely that BC gives rise to the better performance of DC_BC_CC compared to CC.

Figure 7 shows that the precision of all approaches decreases drastically when $k$ goes from 1 to 500. After around 1000 records there is a very low chance of obtaining relevant publications with any of the approaches.

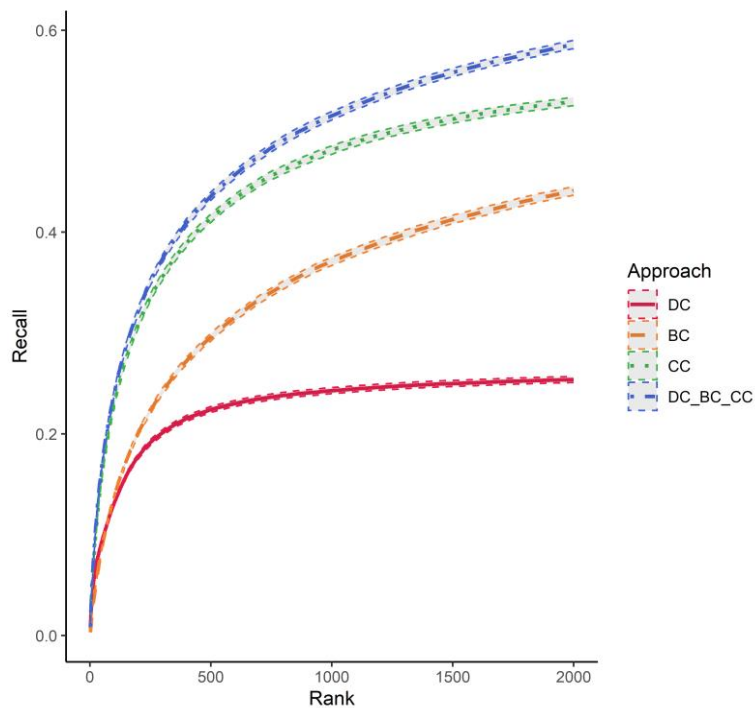

*Figure 6: Recall@k (k = 1, 2, …, 2000) of the approaches with a 95% confidence interval.*

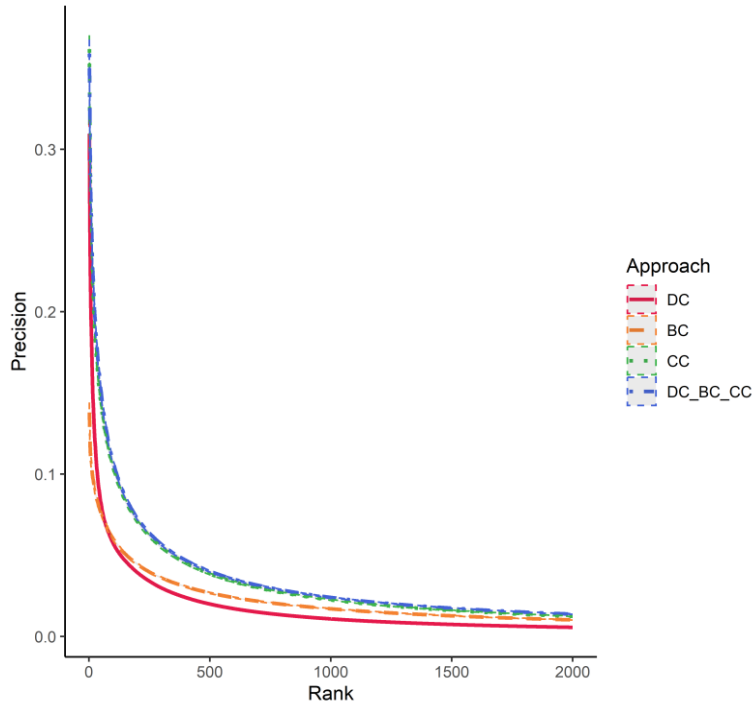

*Figure 7: Precision@k (k = 1, 2, ..., 2000) of the approaches with a 95% confidence interval.*

To assess the plausibility of the results, we made a manual assessment of the top 3 ranked publications by the DC_BC_CC approach that had not resulted in a hit (see GitHub repository). The assessment showed that these publications generally address topics found in the corresponding seeds. It should be noted that the seeds do not always address a coherent topic. Users are likely to select seeds that address the same topic more consistently. A random selection of seeds was chosen to avoid biases towards any of the included relatedness approaches. Nonetheless, the manual assessment indicates that the DC_BC_CC approach in general retrieves topically relevant publications.

## Discussion and conclusions

Using systematic reviews as a baseline, we have compared the performance of DC, CC and BC for seed-based information retrieval. The results clearly show an advantage for CC over BC and DC. As discussed in the section "In recent time, several systems have emerged that use citation-relations for information retrieval, including Inciteful, ResearchRabbit, LitMaps and Connected Papers. The idea to retrieve publications related to seeds by citation relations is not new. Still, there is a lack of studies comparing the performance of different methods. Considering earlier studies on the theme, we further believe that the problem of how to best elaborate seed-based systems needs to be studied more structured. In this study, we focus on the retrieval performance of simple citation-based approaches for seed-based search, believing that our work can form a baseline for future studies that implement more advanced approaches.

To the best of our knowledge, no previous study has compared the performance of DC, BC and CC for seed-based information retrieval. In this paper, we compare the performance of these approaches. In addition, we include the RA score as well as combined approaches in the comparison. In the next section, we report on related research focusing on seed-based approaches. Data are presented in the following section. The section "Methods" treats the citation relations we have used for seed-based information retrieval. This section is followed by the results. Implications for seed-based approaches are discussed in the last section, including recommendations for future research.

Related research", CC has shown to perform well in other studies (Janssens et al., 2020; Janssens & Gwinn, 2015) and correlate with other similarity measures (Colavizza et al., 2018). However, combining the three approaches outperforms the exclusive use of CC in the current study. Aside from this empirical evidence, there are also disadvantages with an exclusive use of CC from a theoretical perspective. CC can only find publications that have already been cited, which makes it impossible to retrieve new publications that have not been cited yet. This is possible using DC or BC, which is a strong argument to include at least one of these measures in applications using citation-based information retrieval.

It should be noted that the results may be biased towards some of the citation approaches, because researchers may have used such approaches in the processes to write the systematic reviews that we have used as baselines. Direct citations and co-citations are most likely to be biased because these approaches are supported by several systems. Bibliographic coupling is less likely to be biased. The reason for this is that bibliographic coupling is not supported by systems easily available to researchers. The extent to which this bias may affect the results is not known, though.

The results also indicate that combining citation-based approaches with textual approaches enhances the performance of seed-based information retrieval. This is in line with previous research (Lu et al., 2007; Steinert et al., 2015). It should be acknowledged that the approach we used to combine the citation-based approaches with RA is rather simplistic and that more sophisticated approaches may show a greater improvement of the performance. The results from this study may guide approaches combining citation-based and textual approaches in their choice of citation similarity measures.

It is possible that the citation network contains relevant publications not reached by DC, BC and CC. The approach suggested by (Liben-Nowell & Kleinberg, 2003) and further developed by (Y. Liang et al., 2011) may find such publications. The study by Liang et al. (2011) indicates that such an approach may perform better than co-citations and bibliographic coupling (named cocoupling by the authors). However, their empirical study is small in scale and combinations of DC, CC and BC were not included.

Another possible direction for improvement of the performance of seed-based approaches is to normalize the citation measures. For example, bibliographic coupling can be normalized to the number of references in the publications as well as to the number of citations to the references of the publications and co-citations can be normalized similarly. Rodriguez-Prieto (2019) suggested and tested one approach for normalization. However, our results show that using more than one citation-relation performs better than using a single approach. Future research may focus on how to normalize these different measures and combine them most effectively.

Several previous studies have used information from in-text citations (Habib & Afzal, 2017, 2019; Ortuño et al., 2013; Tran et al., 2009). Such approaches may enhance retrieval performance by taking into account location of citations and frequency of a particular citation in a paper. However, these approaches are costly in terms of computational power. Furthermore, the availability of fulltext publication data is restricted (nevertheless growing with the increasing number of open access publications). Because of the increased cost to perform such calculations they must considerably enhance retrieval performance to be justified. It is not clear from these studies if and to which extent they enhance performance in comparison with the simplistic approaches used in this study.

In conclusion, the results of this study indicate that CC is a better choice than BC and DC for seed-based information retrieval, but combining the approaches increases the retrieval effectiveness. We suggest that future research use more structured approaches to evaluate methods for seed-based retrieval of publications, including comparative approaches as well as the elaboration of common data sets and baselines for evaluation. The data used in this study is openly available for others to use for future studies on seed-based information retrieval. The present study may give guidance on how to incorporate citation-relations in more sophisticated approaches.

## Data and code availability

Data and code used for data analysis in this study is openly available in GitHub at https://github.com/petersjogarde/papers/tree/main/seed_based_ir.

## Author contributions

Peter Sjögårde: Conceptualization; methodology; software; formal analysis; writing—original draft; writing—review & editing; visualization. Per Ahlgren: Conceptualization; methodology; formal analysis; writing—original draft; writing—review & editing.

## Competing interests

The authors declare no competing interests.

## Funding information

Peter Sjögårde was funded by The Foundation for Promotion and Development of Research at Karolinska Institutet.

---

[i] https://inciteful.xyz/
[ii] https://www.researchrabbit.ai/
[iii] https://www.litmaps.com/
[iv] https://www.connectedpapers.com/